\documentclass[reprint,aps,prl,twocolumn,superscriptaddress,longbiblioography]{revtex4-1}
\usepackage{amssymb,amsmath}
\usepackage{physics}
\usepackage{subfigure}
\usepackage{mathtools}
\usepackage[pdftex]{graphicx,xcolor}
\usepackage[colorlinks=true,linkcolor=blue,urlcolor=blue,citecolor=blue,pdfusetitle,linktocpage]{hyperref}

\begin{document}
\title{Quantum Metrology Protected by Hilbert Space Fragmentation}

\author{Atsuki Yoshinaga}
\email{yoshi9d@iis.u-tokyo.ac.jp}
\affiliation{Department of Physics, The University of Tokyo,
5-1-5 Kashiwanoha, Kashiwa, Chiba 277-8574, Japan}
\affiliation{Research Center for Emerging Computing Technologies, National institute of Advanced Industrial Science and Technology (AIST), Central2, 1-1-1 Umezono, Tsukuba, Ibaraki 305-8568, Japan}

\author{Yuichiro Matsuzaki}
\email{matsuzaki.yuichiro@aist.go.jp}
\affiliation{Research Center for Emerging Computing Technologies, National institute of Advanced Industrial Science and Technology (AIST), Central2, 1-1-1 Umezono, Tsukuba, Ibaraki 305-8568, Japan}

\author{Ryusuke Hamazaki}
\email{ryusuke.hamazaki@riken.jp}
\affiliation{Nonequilibrium Quantum Statistical Mechanics RIKEN Hakubi Research Team, RIKEN Cluster for Pioneering Research (CPR), RIKEN iTHEMS, Wako, Saitama 351-0198, Japan}

\begin{abstract}
We propose an entanglement-enhanced sensing scheme that is robust against spatially inhomogeneous always-on Ising interactions. 
Our strategy is to tailor coherent quantum dynamics employing the Hilbert-space fragmentation (HSF), a recently recognized mechanism that evades thermalization in kinetically constrained many-body systems. 
Specifically, we analytically show that the emergent HSF caused by strong Ising interactions enables us to design a stable state where part of the spins is effectively decoupled from the rest of the system. 
Using the decoupled spins as a probe to measure a transverse field, we demonstrate that the Heisenberg limited sensitivity is achieved without 
suffering from thermalization.
\end{abstract}
\maketitle

\paragraph*{Introduction.---}
Taming entanglement and coherence of a multiple qubit system is a crucial task in today's quantum technology. One of the most notable applications featuring quantum advantage is quantum metrology, where entanglement enables the realization of the enhanced sensitivity in estimating external fields~\cite{giovannetti2004quantum,toth2014quantum,RevQSens2017RevMod}.
For a given number $N$ of probe spins to measure the fields, the uncertainty of the estimation can be reduced in proportion to $N^{-1}$ for entangled states, which is called the Heisenberg limit (HL). In contrast, the corresponding scaling for separable states becomes only $N^{-1/2}$, which is known as the standard quantum limit (SQL).
Due to the fundamental and practical interests, quantum metrology has extensively been studied both theoretically~\cite{yurke19862,wineland1992spin,wineland1994squeezed,huelga1997improvement,lee2002quantum,giovannetti2006quantum,tanaka2015proposed} and experimentally~\cite{jones2009magnetic,lucke2011twin,muessel2014scalable,hosten2016measurement,long2022entanglement,cao2022experimental}.

One major challenge for quantum metrology is to precisely control the dynamics of many-body interacting systems.
On the one hand, interactions among qubits are necessary for preparing entangled states. 
On the other hand, complicated interactions, which are in general spatially inhomogeneous in actual experiments, make the many-body system thermalize.
In fact, recent studies on quantum dynamics elucidate that even isolated system can thermalize due to the eigenstate thermalization hypothesis (ETH) \cite{deutsch1991quantum,srednicki1994chaos,tasaki1998quantum,rigol2008thermalization}, which states that every energy eigenstate becomes locally thermal.
This effect of thermalization \cite{park2016disappearance} would spoil the sensitivity more severely when target magnetic fields become weaker compared with the interactions.

To overcome this unwanted effect of interactions, several approaches have been proposed.
One possible approach is the dynamical decoupling, 
where a sequence of pulses is applied
to eliminate unwanted terms in Hamiltonians \cite{waugh1968approach,haeberlen1968coherent,stollsteimer2001suppression,wocjan2002simulating,zhou2020quantum}.
In general, this method demands
performing a large number of precise pulse operations. 
Another recent approach~\cite{dooley2021robust} that does not involve active operations is to utilize quantum many-body scars~\cite{bernien2017probing,turner2018weak,desaules2022extensive,papic2022weak,dooley2022entanglement}, which are non-thermalizing eigenstates in certain interacting Hamiltonians.
However, this approach is based on Hamiltonians with fine-tuned interactions and hence
susceptible to, e.g., the spatially inhomogeneous perturbations.

Hilbert space fragmentation (HSF) is another novel mechanism that prohibits thermalization in interacting non-integrable systems and has gathered recent attention~\cite{moudgalya2019thermalization,sala2020ergodicity,khemani2020localization,serbyn2021quantum,papic2021weak,moudgalya2021quantum,moudgalya2021hilbert}.
In some models with kinetic constraints,
Hilbert space is fractured into
exponentially many invariant subspaces, which leads to non-ergodicity.
This phenomenon also appears in an effective model that describes the transverse field Ising model (TFIM) in the limit of a weak field \cite{yoshinaga2021emergence, hart2022hilbert,balducci2022localization}. 
In this model, eigenstates can involve ``frozen regions," in which spins in the z direction cannot be dynamically flipped.
The eigenstates with frozen regions appear due to a constraint arising from the emergent conservation of the interaction energy in the weak-field limit and break the ETH and thermalization.
Notably, the structure of the HSF does not rely on the translation invariance and fine-tuning of the Hamiltonian, in stark contrast to typical models hosting quantum many-body scars.

\begin{figure*}[t]
\centering
\includegraphics[width=2\columnwidth]{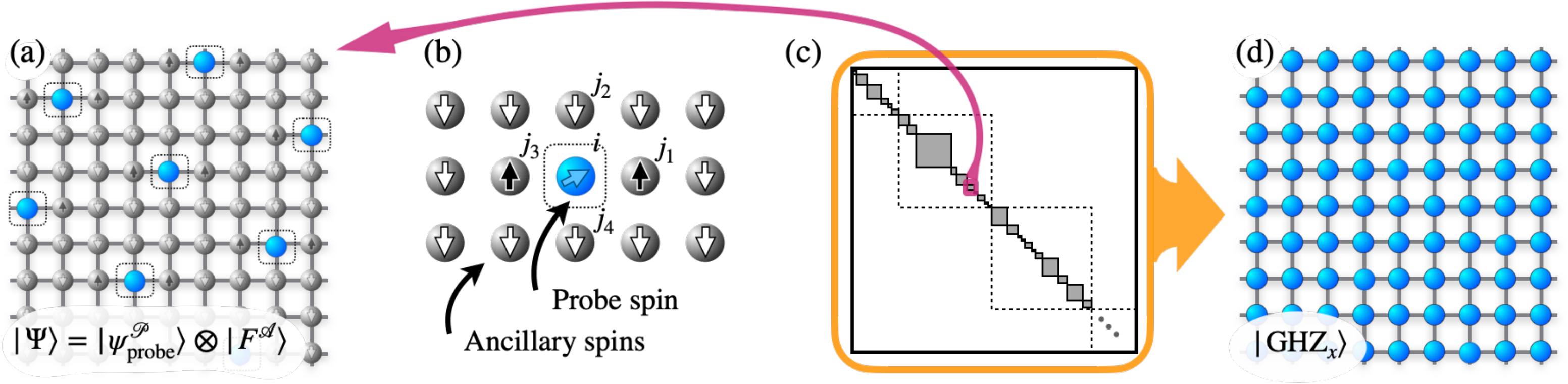}
\caption{
(a) 
Schematic of how we split the system into probe spins and ancillary spins for our quantum sensing scheme.
The blue sites surrounded by the dotted lines represent the probe spins, and the other gray sites correspond to the ancillary spins.
We can achieve the sensitivity with the Heisenberg limit (HL) by using $|\Psi\rangle$ (see Eq.~(\ref{eq:defpsi})) with probe spins being the Greenberger-Horne-Zeilinger (GHZ) state.
(b) Spin configuration around a probe spin in (a), which induces coherent dynamics of the probe spins and dynamical freezing of the ancillary ones.
Each ancillary spin is an eigenstate of $\hat \sigma_i^z$, which corresponds to either spin-up (such as in $j_1$ and $j_3$) or down (such as in $j_2$ and $j_4$) state.
(c) Schematic picture of the emergent Hilbert space fragmentation (HSF) in our transverse-field Ising model with a weak-field limit.
Emergent conservation law of the number of domain walls block-diagonalizes the Hamiltonian, which is further block-diagonalized due to the HSF.
(d) 
Illustration of the GHZ state on a square lattice, which is used as a probe state in the conventional approach.
The GHZ state in (d) corresponds to 
$\left(
\bigotimes_{j=1}^N|+\rangle_{j} + i \bigotimes_{j=1}^N|-\rangle_{j}\right
)/\sqrt2$, which contains a superposition of many computational basis states that spread across the fragmented subspaces.
On the other hand,
the state $|\Psi\rangle$ in (a) corresponds to a superposition of states in a restricted subspace in the fragmented Hilbert space.
}
\label{fig:Fig1}
\end{figure*}

In this Letter, we propose a novel entanglement-enhanced sensing scheme in a strongly interacting inhomogeneous Ising model in two dimension, where the emergent HSF protects the relevant quantum coherence against interactions.
Our strategy is to design a metrologically useful state arranged as in Fig.~\ref{fig:Fig1}~(a,b), where the probe spins are embedded in the ancillary spins.
This state belongs to one of the fragmented subspaces in the TFIM with a weak field limit, which exhibits the emergent HSF as shown in Fig.~\ref{fig:Fig1}~(c), and thus evades thermalization.
More concretely, the probe spins undergo tailored coherent dynamics just with additional bias fields, being decoupled from the ancillary spins that are dynamically frozen.
We analytically show that our scheme reaches the Heisenberg-limited sensitivity in estimating the target transverse field for sufficiently strong interactions.
Our method is robust under various perturbations, such as inhomogeneity, additional longitudinal fields, and certain changes in the lattice structure and spatial dimensions.

\paragraph*{Quantum sensing in an interacting system.---}
We consider a system of spin-$1/2$ particles (qubits) where always-on  Ising interactions exist between them.
We here assume that the spins are arranged in a two-dimensional square lattice, although generalization to higher-dimensions and other types of lattices are straightforward.
The system is exposed to a weak target magnetic field with magnitude $\omega$, which we try to estimate by quantum sensing.
The Hamiltonian is then given by
\begin{align}
\hat H_{\rm TFIM} &= \hat H_\omega + \hat H_{\rm int}, \label{eq:defHall}\\
\hat  H_\omega &= \frac{\omega}2 \sum_{i} \hat \sigma_j^x,\\
\hat H_{\rm int} &= -\sum_{ \langle i,j \rangle } J_{ij} \hat \sigma_i^z \hat \sigma_j^z.
\label{eq:defHint}
\end{align}
where $\langle i,j \rangle$ indicates that the sites $i$ and $j$ are nearest neighbors and we set $\hbar=1$.
Here, $J_{ij} = \bar{J}+\Delta J_{ij}$ denotes the Ising coupling constant, where $\bar{J}$ does not depend on $\langle i,j \rangle$.
We assume that $|\Delta J_{ij}|$ does not exceed $|\bar{J}|/2$, i.e., $\max_{i,j}2|\Delta J_{ij}|/|\bar{J}|=:k<1$.
Without loss of generality, we consider the ferromagnetic case hereafter $(\bar{J}>0)$.

Throughout this Letter, we adopt the Ramsey scheme~\cite{RevQSens2017RevMod} summarized as follows:
(i) prepare initial probe spins in a metrologically useful state;
(ii) let them be exposed to the static target field,
whose Hamiltonian is given by $\hat  H_\omega$, for a duration time $T_{\rm int}$;
(iii) perform a projective measurement described by an operator $\hat P_s$ and obtain an outcome;
and (iv) estimate the value of $\omega$ from the outcomes obtained by the repetition of (i)-(iii).
The uncertainty of the estimation of $\omega$ under this scheme is calculated as
\begin{align}
\delta \omega 
=\frac{\Delta P_s}{\left|\frac{\partial P_s}{\partial \omega}\right|\sqrt{M}},
\label{eq:errorOmegaDef}
\end{align}
where $P_s=\langle \hat P_s \rangle$ denotes the expectation value of $\hat{P}_s$, which corresponds to a probability for the projection onto the desired basis to successfully occur.
Here, $\Delta P_s=\sqrt{P_s(1-P_s)}$ denotes the standard deviation of $\hat P_s$, and $M$ denotes the number of repetitions of the measurements \cite{pezze2018quantum}.
For a total available time $T_{\rm all}$, the number $M$ is calculated as $M=T_{\rm all}/T_{\rm sens}$, where $T_{\rm sens}$ denotes a combined time of the three procedures (i)-(iii) of the sensing scheme.
For simplicity, below we take $T_{\rm sens} = T_{\rm int}$
by assuming that $T_{\rm int}$ for (ii) is much longer than the other duration times for (i) and (iii).

To begin with, let us consider quantum sensing in the absence of the interaction $\hat H_{\rm int}$.
In this case, we can estimate $\omega$ with the HL by preparing the Greenberger-Horne-Zeilinger (GHZ) state $|{\rm GHZ}_x\rangle :=
\left(
\bigotimes_{j=1}^N|+\rangle_{j} + \bigotimes_{j=1}^N|-\rangle_{j}\right
)/\sqrt2
$ as a probe state \cite{greenberger1990bell,MerminGHZ1990} in (i),
where $|\pm\rangle_{j}$ denote eigenstates of $\hat \sigma_j^x$ with eigenvalues $\pm 1$ and $N$ denotes the number of spins.
After this initial state 
acquires the relative phase $\omega N T_{\rm int}$ through (ii), we perform a projective measurement $\hat P_s' = |{\rm GHZ'}_x\rangle\langle {\rm GHZ'}_x|$ with
$
|{\rm GHZ'}_x\rangle :=
\left(
\bigotimes_{j=1}^N|+\rangle_{j} + i \bigotimes_{j=1}^N|-\rangle_{j}\right
)/\sqrt2
$
in (iii), and finally
we estimate $\omega$ from the relation
$
\langle \hat P_s' \rangle = (1/2)(1+ \sin{(\omega N T_{\rm int})}).
$
Throughout this paper we assume that the target field $\omega$ is weak and satisfies $\omega N T_{\rm int}=\mathcal{O}(N^0)\ll 1$ \cite{one_myfoot}. 
We also assume $T_{\rm int} = \mathcal{O}(N^0)$ unless otherwise mentioned.
The uncertainty $\delta\omega$ of the estimation is then calculated from Eq.~(\ref{eq:errorOmegaDef}) as
$\delta \omega = N^{-1} (T_{\rm int}T_{\rm all})^{-1/2}$.
This demonstrates that the HL is achieved by using the GHZ state in the absence of the internal interaction $\hat H_{\rm int}$.
We note that the projective measurement of $\hat P_s'$ can be replaced with a parity measurement described by $(1+\prod_i^{N}\hat \sigma_i^y )/2$ along with an appropriate single spin rotation \cite{two_myfoot}.

\begin{figure}[t]
\centering
\includegraphics[width=1\columnwidth]{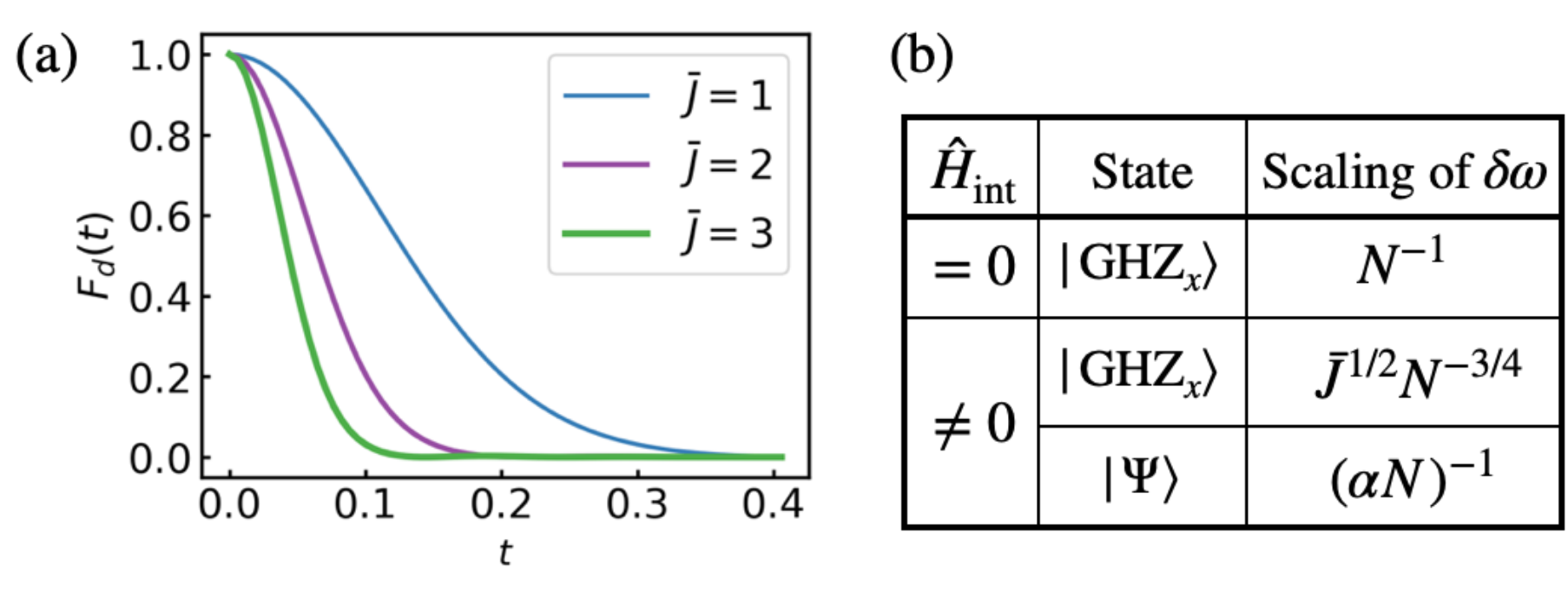}
\caption{
(a) Decay of the dynamical fidelity $F_d(t)$, which compares the time evolutions from the GHZ state $|{\rm GHZ}_x\rangle$ with respect to $\hat H_{\omega}$ and $\hat H_{\rm TFIM}$ for three values of $\bar{J}$. We use a $N=3\times4$ lattice system surrounded by fixed down spins. 
Spatial fluctuations of the interaction $\Delta J_{ij}$ are generated from Gaussian random variables by setting the mean and the variance as zero and $ 0.3\bar{J}$, respectively.
We fix the transverse field $\omega=0.4$ in all of the cases.
(b) 
The asymptotic dependence of $\delta \omega$ on $N$ and the interaction strength $\bar{J}$ for three Ramsey schemes. We compare the sensing schemes using initial states that
are explained in the caption of Fig.~\ref{fig:Fig1}, where we take $|\psi_{\rm probe}^{\mathcal{P}}\rangle= |{\rm GHZ}_x^{\mathcal{P}}\rangle$ for the state $|\Psi\rangle$ here.
Note that the case for $\hat H_{\rm int}\neq0$ with $|{\rm GHZ}_x\rangle$ is achieved by shortening the duration time $T_{\rm int}$ as $N$ or $\bar{J}$ increases, in contrast to the other two cases, where $T_{\rm int}$ is assumed to be a constant.
}
\label{fig:Fig2}
\end{figure}

However, the sensitivity decreases
when $\hat H_{\rm int}$ is taken into consideration.
Due to the flipping of spin states from 
$|\pm \rangle_{i}|\pm \rangle_{j}$ to $|\mp \rangle_{i}|\mp \rangle_{j}$
caused by Ising-type interactions of $\hat H_{\rm int}$, the probe state after (ii) no longer remains in a simple superposition of $\bigotimes_{j=1}^N|+\rangle_{j}$ and $\bigotimes_{j=1}^N|-\rangle_{j}$.
To show the destructive effect of the interaction, we calculate in Fig.~\ref{fig:Fig2}~(a) the time evolution of the dynamical fidelity
$
F_d(t) := \left|\langle {\rm GHZ}_x|e^{ i \hat H_{\omega} t} e^{ - i \hat H_{\rm TFIM} t}|{\rm GHZ}_x\rangle \right|^2
$,
which quantifies the difference between the ideal state evolved by $\hat H_{\omega}$ and the actual state evolved by $\hat H_{\rm TFIM}$ with nonzero interaction
$\hat H_{\rm int}$.
The rapid decay of $F_d(t)$ in Fig.~\ref{fig:Fig2}~(a) implies that the probe state is unstable under the effect of the interaction.
The decay rate increases as the interaction becomes stronger.
This implies that naive sensing with
the GHZ states, as illustrated in Fig.~\ref{fig:Fig1}~(d), will be challenging, especially under the strong always-on Ising interactions.

We note that it is possible to achieve a sensitivity beyond the SQL but below the HL with our model using the GHZ state.
The idea is to appropriately tune the duration time $T_{\rm int}$ in the step (ii) so that the effects from the interaction are minimized.
Specifically, if we decrease $T_{\rm int}$ as $T_{\rm int}=\mathcal{O}(\bar{J}^{-1} N^{-1/2})$ for increasing $N$, the uncertainty of the estimation scales as $\delta\omega=\mathcal{O}(\bar{J}^{1/2}N^{-3/4})$ (see also Supplemental Material, (SM) \cite{supplemental}).
This scaling is called the Zeno scaling \cite{matsuzaki2011magnetic,chin2012quantum}.
While the scaling exceeds the SQL, it is still unsatisfactory since the sensitivity becomes severely worse as the interaction strength becomes stronger.

\paragraph*{HSF-protected quantum metrology.--}
We now illustrate our entanglement-enhanced sensing scheme that is robust against strong always-on-Ising coupling with spatial inhomogeneity.
Instead of using all spins as a probe (see also Fig.~\ref{fig:Fig1}~(d)), we design a state so that a fraction of probe spins are embedded in the ancillary spins as shown in Fig.~\ref{fig:Fig1}~(a).
Specifically, we take the following initial state in step (i):
\begin{align}
|\Psi\rangle := |\psi_{\rm probe}^{\mathcal{P}}\rangle
\otimes
|F^{\mathcal{A}}\rangle\label{eq:defpsi}
\end{align}
Here, 
$|\psi_{\rm probe}^{\mathcal{P}}\rangle$ denotes a state of $\alpha N$ probe spins, and $|F^{\mathcal{A}}\rangle$ denotes that of $(1-\alpha)N$ ancillary spins, where we take a constant $\alpha$ as $\alpha = 1/11$.
The superscript $\mathcal{P}$ ($\mathcal{A}$) indicates that the state is defined on probe (ancillary) spins.
Figure~\ref{fig:Fig1}~(a) illustrates how we divide the system into these two groups of spins.
Each probe spin is interspersed among the ancillary spins.
Figure~\ref{fig:Fig1}~(b) illustrates the spin configuration of the ancillary spins around each probe spin in Fig.~\ref{fig:Fig1}~(a).

Notably, our model exhibits the HSF in the weak-transverse-field limit,
which makes $|\Psi\rangle$ a non-ergodic state with $|F^{\mathcal{A}}\rangle$ being a frozen region and leads to the coherent time-evolution of $|\psi_{\rm probe}^{\mathcal{P}}\rangle$ in step (ii).
Here, ``frozen'' means that the spins cannot be flipped under the time evolution.
In particular, we show that the following approximation holds for any observable $\hat P_s$ with large $\bar{J}/\omega$ 
\begin{align}
\langle\Psi|
e^{i \hat H_{\rm total} t}
\hat P_s
e^{-i \hat H_{\rm total} t}
|\Psi\rangle
\simeq
\langle\Psi|
e^{i \hat H_{\omega}^{\mathcal{P}} t}
\hat P_s
e^{-i \hat H_{\omega}^{\mathcal{P}} t}
|\Psi\rangle \label{eq:decouplePsi},
\end{align}
where $\hat H_{\rm total} = \hat H_{\rm TFIM} + \hat H_{\rm shift}^{\mathcal{P}}$ and
\begin{align}
\hat H_{\omega}^{\mathcal{P}} 
&:= \frac{\omega}2\sum_{i\in \rm probe} \hat \sigma_i^x, \label{eq:defHomegaprobe}\\
\hat H_{\rm shift}^{\mathcal{P}} 
&:= - \sum_{i\in \rm probe} h_i^z \hat \sigma_i^z. \label{eq:defHz}
\end{align}
Here, ``$i\in {\rm probe}$" indicates that the sum is taken over all probe spin sites.
As detailed below, we tune $h_i^z$ in $\hat H_{\rm shift}^{\mathcal{P}}$ so that we can cancel out effective longitudinal fields on the probe spins that arise due to $\hat H_{\rm int}$.
Equation (\ref{eq:decouplePsi}) suggests that, for $|\Psi\rangle$, the probe spins are decoupled from the rest of the interacting but dynamically frozen spins and exposed only to the target field $\hat H_{\omega}^{\mathcal{P}}$.

To understand Eq.~(\ref{eq:decouplePsi}), we first note that a spin flip by $\hat{H}_\omega$ with small $\omega$ is suppressed when the flip causes a large change in the energy due to $\hat{H}_{\rm int}$. 
For simplicity, let us start from the case with $\Delta J_{ij}=h_i^z=0$ and $\omega/\bar{J}\rightarrow 0$.
In this case, the large interaction $\hat{H}_{\rm int}$ leads to a constraint where a spin can flip only when two surrounding spins are up and the other two surrounding spins are down.
This constraint results in the occurrence of the HSF as studied in Refs.~\cite{yoshinaga2021emergence, hart2022hilbert}; the effective Hamiltonian has a  block-diagonal structure by the emergent conservation law of the domain-wall (DW) number $\hat n_{DW} := \sum_{ \langle i,j \rangle }  (1-\hat \sigma_i^z \hat \sigma_j^z)/2$, 
which is further fragmented nontrivially as shown in Fig.~\ref{fig:Fig1}~(c).
This suggests  non-ergodicity even within each DW sector.

We next argue that a similar HSF emerges for $\Delta J_{ij}\neq 0$ and that $|\Psi\rangle$ 
corresponds to a state in one of the fragmented subspaces. 
More concretely, $|F^{\mathcal{A}}\rangle$ constitutes a frozen region:
from the construction given in Fig.~\ref{fig:Fig1}~(b), every ancillary spin is always surrounded by at least three down spins.
Then, the action of $\hat{H}_\omega$ is energetically suppressed on this region even for nonzero $\Delta J_{ij}$, since the magnitude of the fluctuation $\Delta J_{ij}$
is assumed not to exceed $\bar{J}/2$. Thus, $|F^{\mathcal{A}}\rangle$ becomes dynamically stable in $\omega/\bar{J}\rightarrow 0$ limit, independent of the state of the probe spins $|\psi_{\rm probe}^{\mathcal{P}}\rangle$.
Due to the constraint, such a frozen region appears in other configurations as well, leading to exponentially many invariant subspaces, which means the occurrence of the HSF.
Our designed initial state $\ket{\Psi}$ then belongs to one of such subspaces and time-evolves only within it (see SM~\cite{supplemental}).

We now discuss the origin of $\hat{H}_{\omega}^{\mathcal{P}}$ in Eq.~(\ref{eq:decouplePsi}), focusing on probe spins. 
Since each probe spin is surrounded by two up and two down frozen spins, the probe spin is effectively exposed to
an effective longitudinal magnetic field $\tilde h_i^z = - \Delta J_{ij_1} + \Delta J_{ij_2} - \Delta J_{ij_3} + \Delta J_{ij_4}$, see Fig.~\ref{fig:Fig1}~(b).
Assuming that each $\tilde h_i^z$
is known from calibration, we can cancel the effective field by choosing ${h}_i^z=-\tilde{h}_i^z$ in Eq.~(\ref{eq:defHz}).
Therefore, $\hat H_{\rm total}$ acting on our state $|\Psi\rangle$ is reduced to $\hat H_{\omega}^{\mathcal{P}}$ when $\omega/\bar{J}\rightarrow 0$.

In our scheme, we perform the Ramsey sensing (i)--(iv) with the following two modifications.
First, we only use the probe spins as a resource of metrology and make the other spins ancillary.
In other words, we prepare $|\Psi\rangle$ with $|\psi_{\rm probe}^{\mathcal{P}}\rangle= |{\rm GHZ}_x^{\mathcal{P}}\rangle$ in (i) and readout outcomes by using a projective operator $\hat P_s=|{\rm GHZ'}_x^{\mathcal{P}}\rangle\langle{\rm GHZ'}_x^{\mathcal{P}}|\otimes \hat I^{\mathcal{A}}$ in (iii).
Second, we additionally apply the shift field $\hat H_{\rm shift}^{\mathcal{P}}$ to the probe spins during the exposure (ii), as discussed in the previous paragraph.
In the limit of $\omega/\bar{J}\rightarrow0$, Eq.~(\ref{eq:decouplePsi}) is exact, and
the uncertainty is calculated as
$\delta \omega = (\alpha N)^{-1}(T_{\rm int}T_{\rm all})^{-1/2}$, which demonstrates the Heisenberg-limited sensitivity.
Our scheme does not require turning off the interactions or controlling $T_{\rm int}$ during the interrogation process.
The table in Fig.~\ref{fig:Fig2}~(b) summarizes three schemes that we introduced in this Letter.
The protocols using the state $|{\rm GHZ}_x\rangle$ suffer from the interactions, while our protocol using $|\Psi\rangle$ achieves the HL for the estimation error for sufficiently large $\bar{J}/\omega$.

\paragraph*{Stability for finite $\omega /\bar{J}$.--}
While the freezing of the ancillary spins discussed above is exact only for $\omega/\bar{J}\rightarrow 0$, we here analytically show that the HL is still achieved in our scheme even for sufficiently small but finite $\omega/\bar{J}$.
To see this, we first evaluate the uncertainty of $\omega$ by taking account of the deviation $\epsilon(t)$ from the approximation in Eq.~(\ref{eq:decouplePsi}) (see SM~\cite{supplemental} for the derivation):
\begin{align}
\delta \omega =
\frac{1}{aNT_{\rm int}}\left(\frac{\langle \hat P_s \rangle_{\rm actual}(1-\langle \hat P_s \rangle_{\rm actual})}{M} + |\epsilon(T_{\rm int})|^2\right)^{1/2},\label{eq:uncertaintywithepsilon}
\end{align}
where $\epsilon(t):=\langle \hat P_s \rangle_{\rm actual}-\langle \hat P_s \rangle_{\rm eff}$ denotes
the difference between $\langle \hat P_s \rangle_{\rm actual}:=\langle \Psi | e^{i \hat H_{\rm total} t} \hat P_s e^{-i \hat H_{\rm total} t} |\Psi\rangle$ and $\langle \hat P_s \rangle_{\rm eff}:=\langle \Psi | e^{i\hat H_{\omega}^{\mathcal{P}} t} \hat P_s e^{-i\hat H_{\omega}^{\mathcal{P}} t} |\Psi\rangle$.
When $\epsilon(T_{\rm int}) = \mathcal{O}(N^0)$, the uncertainty $\delta \omega$ scales as $\mathcal{O}(N^{-1})$ and the HL remains to be achieved.

Now, we can analytically
show that $\epsilon(T_{\rm int}) = \mathcal{O}(N^0)$ from the following inequality \cite{supplemental}:
\begin{align}
|\epsilon(T_{\rm int})|
\leq
\frac{2N\omega}{J_g}
+
2\left(e^{{N\omega}/{J_g}}-1\right)N\omega T_{\rm int},
\label{eq:epsilonbound}
\end{align}
where $J_g = \min_{i}
\left[ 4\bar{J} - \sum_{j\in \langle i,j \rangle} |2 \Delta J_{ij}|\right]\geq 4(1-k) \bar{J}>0$ is evaluated from the minimum energy change associated with a flipping of ancillary spins (remind the assumption $\max_{i,j}2|\Delta J_{ij}|/\bar{J}=k<1$).
Since $N\omega T_{\rm int}=\mathcal{O}(N^0)\ll1$ and $T_{\rm int}=\mathcal{O}(N^0)$ are assumed here as a typical setting in Ramsey-type sensing with GHZ states \cite{degen2017quantum}, $\epsilon(t) = \mathcal{O}(N^0)$ holds.
Furthermore, the deviation becomes $|\epsilon(T_{\rm int})|\ll1$ for $N\omega \ll J_g$, which shows that strong interaction is beneficial
in our scheme.
We note that the bound in Eq.~(\ref{eq:epsilonbound}) is derived by generalizing the error bound discussed in Refs.~\cite{gong2020universal,gong2020error,gong_bounds2022}.
Equation \eqref{eq:epsilonbound} also shows that the effective description of the dynamics becomes valid for the intermediate timescale for a weak target transverse field.
That is, our sensing scheme exploits the HSF that emerges in a prethermal regime \cite{abanin2017rigorous,mori2018thermalization} before evolving into the final equilibrium.

\paragraph*{Discussion.--}
Our scheme leads to better sensitivity for stronger interactions, in stark contrast to conventional methods as summarized in Fig.~\ref{fig:Fig2}~(b).
Importantly, our scheme is robust against the inhomogeneity of the interaction.
The mechanism of the approximate freezing is also applicable for finite-range farther-neighbor interactions, cubic or triangular lattices, as well as the additional presence of weak longitudinal fields.
This is due to the broad applicability of the mechanism of the suppression of spin flips under a weak transverse field and strong Ising interactions.
Therefore, our HSF-protected sensing scheme can be generalized for these situations.

Finally, we describe a possible procedure for creating the entangled state $|{\rm GHZ}_x^{\mathcal{P}}\rangle \otimes |F^{\mathcal{A}}\rangle$ as follows.
We first prepare the GHZ state $|{\rm GHZ}_x\rangle$ using the entire spins,
by, e.g., adiabatically transforming a trivial state into the state $|{\rm GHZ}_z\rangle$ as suggested in Refs.~\cite{choi2017quantum,hatomura2022quantum,matsuzaki2022generation}, and then rotating every spin by the angle $\pi/2$.
Note that the state $|{\rm GHZ}_z\rangle$ corresponds to a superposition of the two ground states of the system Hamiltonian $\hat H_{\rm int}$ in the ferromagnetic case.
Then we obtain our desired state after performing the projection $\hat P_{\Psi} = \mathbb{\hat{I}}^{\mathcal{P}}\otimes |F^{\mathcal{A}}\rangle\langle F^{\mathcal{A}}|$ to $|{\rm GHZ}_x\rangle$, which is equivalent to 
measurement feedback control on the ancillary spins: measuring in the $z$ basis and then applying single-spin rotations depending on the measurement results.

\paragraph*{Conclusion.--}
In this Letter, we have proposed a  quantum sensing scheme
for a  system with spatially non-uniform always-on Ising interactions.
Specifically, we show that 
the Heisenberg limited sensitivity is robustly achieved by designing a tailored state that evades thermalization due to the emergent Hilbert-space fragmentation (HSF).
In this state, the entangled probe spins are decoupled from the rest of the system.
This decoupling is due to a kinetic constraint that approximately emerges in the prethermal regime for strong Ising couplings and allows us to measure a transverse field stably.
Our scheme establishes a novel approach to realize quantum sensing in a quantum many-body system with spatial inhomogeneity by using no dynamical controls.
It is rigidly applicable even when the lattice shape and spatial dimensions are altered, as long as the HSF structure offers us a way to control coherent dynamics without thermalization.

Here, we have introduced a concept of designing quantum states that avoid many-body thermalization by the HSF. 
Beyond quantum metrology, this HSF-protected manipulation of quantum dynamics would be advantageous for other quantum technologies as well, where retaining entanglement in the presence of interactions is crucial.

\begin{acknowledgments}
We thank Zongping Gong for helpful comments on the error bound in constrained dynamics.
This work was supported by Leading Initiative for Excellent Young Researchers MEXT Japan and JST presto (Grant No. JPMJPR1919) Japan.
R.H. was supported by JST ERATO-FS Grant Number JPMJER2204, Japan.
\end{acknowledgments}

\bibliographystyle{apsrev4-1}
\bibliography{BibHSFSensing}

\begin{thebibliography}{67}
\bibitem [{two()}]{two_myfoot}%
  \BibitemOpen
  \href@noop {} {}\bibinfo {note} {The expression $\langle \hat P' \rangle = (1
  + \sin{( \omega N T_{\rm int})})/2$ can also be obtained, for example, when
  we replace $P'$ with a parity measurement (1+$\prod_{i=1}^{N}\hat \sigma_i^y
  )/2$ on probe spins after applying $\pi (l_N -1)/4$ rotation along the $x$
  axis on one of the probe spins, where the integer $l_n$ denotes the reminder
  obtained by dividing $N$ by $4$.}\BibitemShut {Stop}%
\end{thebibliography}

\end{document}


\title{Supplemental Material for ``Quantum Metrology Protected by Hilbert Space Fragmentation''}

\author{Atsuki Yoshinaga}
\email{yoshi9d@iis.u-tokyo.ac.jp}
\affiliation{Department of Physics, The University of Tokyo,
5-1-5 Kashiwanoha, Kashiwa, Chiba 277-8574, Japan}
\affiliation{Research Center for Emerging Computing Technologies, National institute of Advanced Industrial Science and Technology (AIST), Central2, 1-1-1 Umezono, Tsukuba, Ibaraki 305-8568, Japan}

\author{Yuichiro Matsuzaki}
\email{matsuzaki.yuichiro@aist.go.jp}
\affiliation{Research Center for Emerging Computing Technologies, National institute of Advanced Industrial Science and Technology (AIST), Central2, 1-1-1 Umezono, Tsukuba, Ibaraki 305-8568, Japan}

\author{Ryusuke Hamazaki}
\email{ryusuke.hamazaki@riken.jp}
\affiliation{Nonequilibrium Quantum Statistical Mechanics RIKEN Hakubi Research Team, RIKEN Cluster for Pioneering Research (CPR), RIKEN iTHEMS, Wako, Saitama 351-0198, Japan}

\maketitle

\renewcommand{\theequation}{S\arabic{equation}}
\renewcommand{\thefigure}{S\arabic{figure}}
\renewcommand{\thetable}{S\arabic{table}}

\section{Zeno scaling}\label{app:zeno}
We discuss a way to obtain the sensitivity better than the standard quantum limit (SQL) in our system just by controlling the duration time $T_{\rm int}$, which turns out to be below the Heisenberg limit (HL).
Specifically, we show that, by 
sufficiently shortening
$T_{\rm int}$, we can suppress the effect of the residual interaction $\hat H_{\rm int}$ and obtain the Zeno scaling $\delta \omega \propto N^{-3/4}$ \cite{matsuzaki2011magnetic,chin2012quantum,tanaka2015proposed}.
Although this scaling of $N^{-3/4}$ is worse than the HL, it is better than the SQL.
We consider preparing a state $|{\rm GHZ}_x\rangle$ and let it evolve for a time $T_\mathrm{int}$ according to $\hat H_{\rm TFIM}$ in [Eq.~(1) in the main text].
Assuming that $T_\mathrm{int}$ is sufficiently small and that the perturbation series converge, the expectation value of the projective measurement $\hat P_s$ in this case is evaluated as (up to the second order)
\begin{align}
P_s
= \frac{1}{2} + \frac{1}{2} \omega N T_{\rm int} -\frac12 T_{\rm int}^2\sum_{\langle i,j\rangle } J_{ij}^2 + \mathcal{O}(T_{\rm int}^3).
\label{eq:supp A estimP}
\end{align}
Note that we can approximate $\frac12 \sum_{\langle i,j\rangle } J_{ij}^2\sim \bar{J}^2 N$ for simplicity.
The expression (\ref{eq:supp A estimP}) shows that the expectation value of the measurement outcomes is affected by the presence of the interaction term $\hat H_{\rm int}$.
Importantly, as this additional term due to the interaction increases quadratically in time $T_{\rm int}$,
the standard deviation $\sqrt{P_s (1- P_s)}$ increases for large $T_\mathrm{int}$, which contributes to the deterioration of the sensitivity.
On the other hand, as we increase the duration time, $|\partial P_s/\partial \omega |$ becomes large, which contributes to the enhancement of the sensitivity.
These suggest that there is an optimal duration time $T_{\rm int}$ for the sensitivity $\delta \omega$.

Now we evaluate the optimal sensitivity by shortening the duration time $T_{\rm int}$.
Let us take $T_{\rm int}=\tau N^{-1/2-\beta} \bar{J}^{-1-\gamma}$, where $\tau$, $\beta$ and $\gamma$ are constants with $\tau\ll1$.
First, we can see that if either $\beta$ or $\gamma$ is negative,
the second-order term 
in Eq.~(\ref{eq:supp A estimP})
will diverge as we take either $N\rightarrow \infty$ or  $\bar{J}\rightarrow \infty$. 
Thus, we take $\beta,\gamma\geq0$ below.
To make the situation similar to the main text, we again assume that $\omega=\mathcal{O}(N^{-1})$ and set $\omega = \omega_0 N^{-1}$, where $\omega_0$ is a constant.
Then, we can evaluate the uncertainty using Eq.~(4) in the main text and Eq.~(\ref{eq:supp A estimP}) as follows:
\begin{align}
\delta \omega = 
\frac{2}{T_{\rm all}^{1/2}}
\left(
\frac14 \tau^{-1}\bar{J}^{1-\gamma} N^{-3/2-\beta}
- \frac14 \omega_0^2 \tau\bar{J}^{-1+\gamma} N^{-5/2+\beta} 
+ \omega_0 \tau^2 \bar{J}^{2\gamma} N^{-2+2\beta}
- \bar{J}^{1+3\gamma} \tau^3 N^{-3/2+3\beta}
\right)^{1/2}.\label{eq:supp A deltaomega1}
\end{align}
For 
large $N$, this is minimized when we take $\beta=0$.
In this case, $\gamma=0$ gives the optimal sensitivity.
Noting that the first term in the parenthesis dominates in Eq.~(\ref{eq:supp A deltaomega1}), we can estimate the optimal scaling of the uncertainty as
\begin{align}
\delta \omega &\simeq
(\bar{J}/\tau T_{\rm all})^{1/2} N^{-3/4}
=\mathcal{O}(N^{-3/4}),
\end{align}
which demonstrates the Zeno scaling.
Note that the uncertainty increases as $\bar{J}$ becomes larger in this case.
This is consistent with our observation in the main text, which shows a faster decay of the dynamical fidelity for a larger $\bar{J}$ in case of using the state $|{\rm GHZ}_x\rangle$ as the initial state.

\section{Derivation of the uncertainty Eq.~(9) in the main text}\label{app:uncertainty}
We evaluate the uncertainty $\delta \omega$ by taking into account 
the deviation in the approximation Eq.~(6) in the main text with  finite $\omega/\bar{J}$.
For this purpose, let us remind that $\omega$ in our scheme is estimated from a sequence of measurement outcomes.
Suppose that we obtain $M$ outcomes by repeating the step (i)--(iii) using our scheme.
Let us define $\{m_1,m_2,\cdots, m_M \}$ with $m_j \in \{0,1\}$ as the sequence of the measurement outcomes and $S_M = (1/M)\sum_{j=1}^{M} m_j$ as the average of $m_j$.
We estimate the unknown parameter $\omega$ from $S_M$ as follows.
When $M$ goes to infinity, the average $S_M$ is given by the quantum expectation value of the measurement $\langle \hat P_s \rangle_{\rm actual} := \langle \Psi | e^{i\hat H_{\rm total} T_{\rm int}} \hat P_s e^{-i\hat H_{\rm total} T_{\rm int}} |\Psi\rangle$.
Although calculating the $\omega$-dependence of $\langle \hat P_s \rangle_{\rm actual}$ analytically is difficult in general,
we have the approximation Eq.~(6) in the main text, which can be recast as an equation as follows:
\begin{align}
\langle \hat P_s \rangle_{\rm actual} = (1/2)\left(1 +\sin{( \alpha N\omega T_{\rm int})}\right)+
\epsilon(T_{\rm int})
,\label{appeq:prealandomega}
\end{align}
where $\epsilon (T_{\rm int}) := \langle \hat P_s \rangle_{\rm actual} - \langle \hat P_s \rangle_{\rm eff}$ is the deviation in the approximation due to finite $\omega/\bar{J}$ and we substitute $T_{\rm int}$ for $t$ in Eq.~(6).
As discussed in the main text, when $N\omega\ll \bar{J}$, $\epsilon(T_{\rm int})$ becomes small. 
Assuming $N\omega\ll T_\mathrm{int}^{-1}$ as well, we can approximate Eq.~(\ref{appeq:prealandomega}) as $\langle \hat P_s \rangle_{\rm actual} \simeq (1/2)\left(1 + \alpha N\omega T_{\rm int} \right)$.
Let us introduce $ \omega_{M}^{est} := (1/\alpha NT_{\rm int})(2S_M - 1)$, where $\omega_{M}^{est}$ denotes the estimated value of $\omega$ in our scheme.
Importantly, in the limit of large $M$ and $\bar{J}$, we obtain $\omega =\omega_{M}^{est}$.
On the other hand, $\omega_{M}^{est}$ does not necessarily coincide with the actual $\omega$ when $\epsilon (t)$ or $M$ is finite.
Below we consider this case and evaluate the uncertainty.
Assuming $N\omega T_{\rm int}\ll1$, we obtain $\omega \simeq \left(2\langle \hat P_s \rangle_{\rm actual} -1- 2\epsilon(T_{\rm int})\right)/(\alpha N T_{\rm int})$ from Eq.~(\ref{appeq:prealandomega}), and then the uncertainty is calculated as \cite{sugiyama2015precision,takeuchi2019quantum}
\begin{align}
\delta \omega^2
&:= \langle (\omega_{M}^{est} - \omega)^2\rangle_{\rm stat}\\
&=
\left\langle \left(\frac{2S_M - 1}{\alpha
NT_{\rm int}} - \frac{2\langle \hat P_s \rangle_{\rm actual} -1- 2\epsilon(T_{\rm int})}{\alpha
NT_{\rm int}}\right)^2\right\rangle_{\rm stat}\\
&=
\frac{4}{\alpha ^2
N^2 T_{\rm int}^2}\left\langle\left(S_M - \langle \hat P_s \rangle_{\rm actual}\right)^2 +  \epsilon(T_{\rm int})^2\right\rangle_{\rm stat}\\
&=
\frac{4}{\alpha ^2N^2 T_{\rm int}^2}\left(\frac{\langle \hat P_s \rangle_{\rm actual}(1-\langle \hat P_s \rangle_{\rm actual})}{M} + \epsilon(T_{\rm int})^2\right),
\end{align}
where $\langle\cdot\rangle_{\rm stat}$ is the statistical average of the outcomes $m_j$.
In the last line, we used  $\left\langle m_j \right\rangle_{\rm stat} = \langle \hat P_s \rangle_{\rm actual}$ and $\left\langle m_i m_j \right\rangle_{\rm stat} = 0$ for $i\neq j$ to derive 
\begin{align}
\left\langle\left(S_M - \langle \hat P_s \rangle_{\rm actual}\right)^2\right\rangle_{\rm stat} 
&=
\left\langle 
\sum_{i}\sum_{j} \left(m_i - \langle \hat P_s \rangle_{\rm actual}\right)\left(m_j - \langle \hat P_s \rangle_{\rm actual}\right)
\right\rangle_{\rm stat}/M^2\\
&=
\left\langle 
(1/M)\sum_{i} \left(m_i - \langle \hat P_s \rangle_{\rm actual}\right)^2
\right\rangle_{\rm stat}/M\\
&=
\langle \hat P_s \rangle_{\rm actual}(1-\langle \hat P_s \rangle_{\rm actual})/M.
\end{align}
This evaluation means that $\delta\omega$ consists of the statistical error coming from finite $M$ and the systematic error coming from finite $\epsilon(T_{\rm int})$.

\section{Derivation of the error bound Eq.~(10) in the main text}\label{app:bound}
Here we show how one can derive the upper bound given in Eq.~(10) in the main text.
We introduce the following theorem, which is a generalized version of the bound in Ref.~\cite{gong2020error}.

\begin{figure}[t]
\centering
\includegraphics[width=0.4\columnwidth]{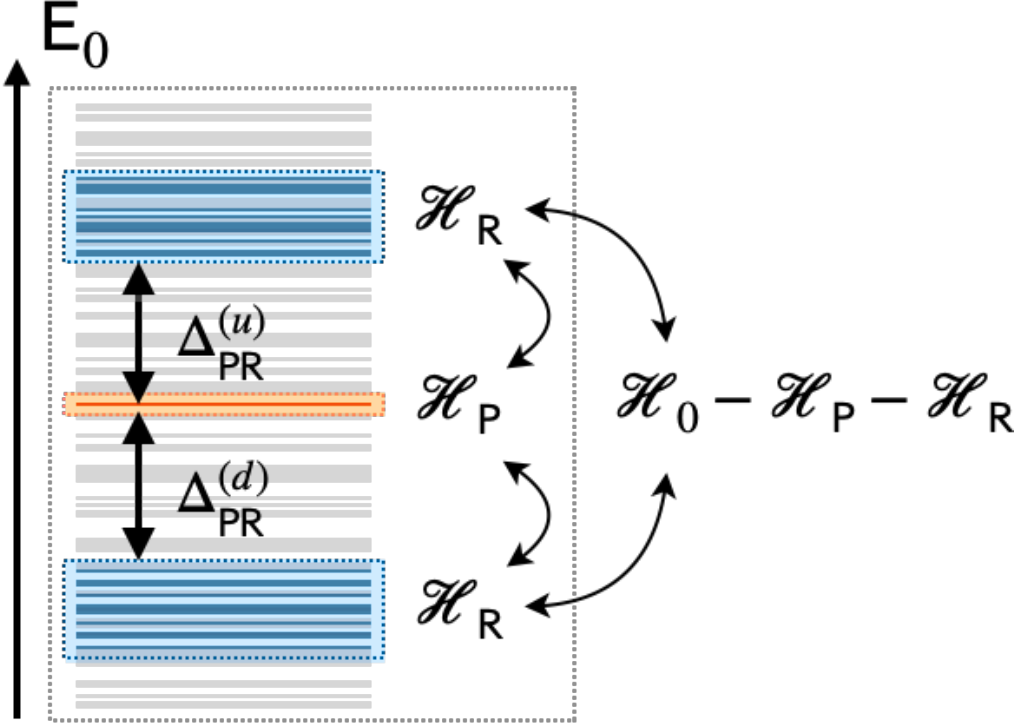}
\caption{
Energy spectrum of $\hat {\mathsf H}_0$.
The energy eigenspace $\mathcal{H}_\mathsf{P}$ does not need to be isolated from the rest of the spectrum with an energy gap.
Another eigenspace $\mathcal{H}_\mathsf{R}  \subseteq \mathcal{H}_{0} \setminus \mathcal{H}_\mathsf{P}$ is connected via $\hat {\mathsf V}$ from $\mathcal{H}_\mathsf{P}$, where $\mathcal{H}_0$ denotes the total Hilbert space of $\hat {\mathsf H}_0$.
We assume that $\mathcal{H}_\mathsf{P}$ and $\mathcal{H}_\mathsf{R}$ are separated by a finite energy gap $\Delta_{\mathsf P\mathsf R}\equiv \min \{\Delta_{\mathsf P\mathsf R}^{(u)}, \Delta_{\mathsf P\mathsf R}^{(d)}\}$.
}
\label{fig:FigSupBoundproof}
\end{figure}

\begin{mythm}[Universal error bound]
Consider a Hamiltonian $\hat{\mathsf H}=\hat{\mathsf H}_0+\hat{\mathsf V}$ of a quantum system, where $\hat {\mathsf H}_{0}$ is regarded as a non-perturbed part and $\hat{\mathsf V}$ is a small perturbation.
Let $\mathcal{H}_\mathsf{P}$ be a certain energy eigenspace of $\hat {\mathsf H}_{0}$ and write $\hat {\mathsf P}$ as the projection operator on $\mathcal{H}_\mathsf{P}$.
Suppose that another energy eigenspace $\mathcal{H}_\mathsf{R}$ ($\mathcal{H}_\mathsf{P}\cap \mathcal{H}_\mathsf{R} = \phi$) of the Hamiltonian $\hat {\mathsf H}_{0}$ satisfies the followings [see Fig.~\ref{fig:FigSupBoundproof}];
(I)
$
\hat {\mathsf P} \hat {\mathsf V} (1- \hat {\mathsf P}) + (1- \hat {\mathsf P}) \hat {\mathsf V} \hat {\mathsf P} 
=
\hat {\mathsf P} \hat {\mathsf V} \hat {\mathsf R} + \hat {\mathsf R} \hat {\mathsf V} \hat {\mathsf P}
$, where $\hat {\mathsf R}$ is the projection onto the energy eigenspace $\mathcal{H}_\mathsf{R}$;
(II)
Energy spectra of $\mathcal{H}_\mathsf{P}$ and $\mathcal{H}_\mathsf{R}$ are separated by a finite energy gap $\Delta_{\mathsf P\mathsf R}>0$, where $\Delta_{\mathsf P\mathsf R}:=\min_{|E\rangle \in \mathcal{H}_\mathsf{P}, |E’\rangle \in \mathcal{H}_\mathsf{R}} |E-E’|$.
Then, starting from an initial state belonging to $\mathcal{H}_\mathsf{P}$, we have the following error bound between the original dynamics described by $\hat{\mathsf H}$ and effective dynamics by $\mathsf{\hat{H}}_{\mathsf{P}}=\hat{\mathsf{P}}\hat{\mathsf H}\hat{\mathsf{P}}$:
\begin{align}
\epsilon(t)
:=&
\left\| \hat {\mathsf P} \left(
e^{i\hat {\mathsf H} t} \hat {\mathsf O} e^{- i \hat {\mathsf H} t}
-
e^{i\hat {\mathsf H}_{{\mathsf P}} t} \hat {\mathsf O} e^{- i \hat {\mathsf H}_{{\mathsf P}} t}
\right) \hat {\mathsf P} \right\|\\
\leq&
\frac{4\|\hat {\mathsf V}\|}{\Delta_{\mathsf P\mathsf R}}
+
2
\left(e^{2\|\hat {\mathsf V}\| / \Delta_{\mathsf P\mathsf R}}-1\right)\|\hat {\mathsf V}\|t
\end{align}
for any $\hat {\mathsf O}$ with $\|\hat {\mathsf O}\|=1$, where $\|\cdot\|$ denotes the operator norm.
\end{mythm}

In contrast to the bounds in Refs.~\cite{gong2020universal,gong2020error} that rely on the energy gap between $\mathcal{H}_\mathsf{P}$ and the rest of the spectrum, our theorem does not require that the eigenspace $\mathcal{H}_\mathsf{P}$ is isolated in the energy spectrum of $\hat {\mathsf H}_0$, see Fig.~\ref{fig:FigSupBoundproof}.
That means, when $\hat {\mathsf V}$ and $\hat {\mathsf P}$ satisfy the conditions (I) and (II), $\epsilon (t)$ is still bounded by using the energy gap between $\mathcal{H}_\mathsf{P}$ and a part of the spectrum $\mathcal{H}_\mathsf{R}$, which is connected from $\mathcal{H}_\mathsf{P}$ via $\hat {\mathsf V}$.
In fact, the upper bound in Ref.~\cite{gong2020error} corresponds to the special case at which $\hat {\mathsf R} = 1-\hat {\mathsf P}$.

Our theorem can be derived just by following the proof in Ref.~\cite{gong2020error} with a slight modification.
The proof is divided into two parts.
In the first step, we derive
\begin{align}
\epsilon(t)\leq
4\|\hat {\mathsf T}\|
+
2(e^{2\|\hat {\mathsf T}\|}-1)
\|\hat {\mathsf V}\|t.
\end{align}
Here, 
$\hat {\mathsf T}$ is an anti-Hermitian operator $\hat {\mathsf T}$ (i.e., $\hat {\mathsf T}^\dagger = -\hat {\mathsf T}$) that satisfies
\begin{align}
[\hat {\mathsf T},\,\hat {\mathsf H}_0] = - \hat {\mathsf V}_{\rm off}
\label{eq:sup:sylvestereq1}
\end{align}
with $\hat {\mathsf V}_{\rm off} 
:=
\hat {\mathsf P} \hat {\mathsf V} (1- \hat {\mathsf P}) + (1- \hat {\mathsf P}) \hat {\mathsf V} \hat {\mathsf P} $. The derivation is obtained and explained in Section ${\rm I\hspace{-.01em}I}$-${\rm B}$ in Ref.~\cite{gong2020error} (see especially Eqs.~(23) and (30)).
In the second step, which is different from Ref.~\cite{gong2020error}, we
derive an upper bound on $\|\hat {\mathsf T}\|$ as follows:
\begin{align}
\|\hat {\mathsf T}\| =
\|{\bar{\mathsf T}}_{\mathsf P\mathsf R}\| 
\leq
\frac{
\|\bar{\mathsf V}_{\mathsf P\mathsf R}\|
}{
\Delta_{\mathsf P\mathsf R}}
\leq
\frac{
\|\hat {\mathsf V}\|
}{
\Delta_{\mathsf P\mathsf R}}
\label{eq:sup:sylvesterinequality}.
\end{align}
Here, 
the matrix ${\bar{\mathsf T}}_{\mathsf P\mathsf R}$ is the solution for the Sylvester equation
\begin{align}
{\bar{\mathsf H}}_{\mathsf 0\mathsf P} {\bar{\mathsf T}}_{\mathsf P\mathsf R}
-
{\bar{\mathsf T}}_{\mathsf P\mathsf R} {\bar{\mathsf H}}_{\mathsf 0\mathsf R}
=
{{\bar{\mathsf V}}}_{\mathsf P\mathsf R},
\label{eq:sup:sylvestereq2}
\end{align}
where 
${{\bar{\mathsf V}}}_{\mathsf P\mathsf R} \equiv \hat {\mathsf P}\hat {\mathsf V}\hat {\mathsf R}$,
${\bar{\mathsf T}}_{\mathsf P\mathsf R} \equiv \hat {\mathsf P}\hat {\mathsf T}\hat {\mathsf R}$,
${\bar{\mathsf H}}_{\mathsf 0\mathsf P} \equiv \hat {\mathsf P}\hat {\mathsf H}_0\hat {\mathsf P}$, and
${\bar{\mathsf H}}_{\mathsf 0\mathsf R} \equiv \hat {\mathsf R}\hat {\mathsf H}_0\hat {\mathsf R}$.
Our observation here is that we can take a solution $\hat {\mathsf T}$ of Eq.~(\ref{eq:sup:sylvestereq1}) as $\hat {\mathsf T} = \hat {\mathsf P}\hat {\mathsf T}\hat {\mathsf R}+\hat {\mathsf R}\hat {\mathsf T}\hat {\mathsf P}$ under the condition (I).
In this case, Eq.~(\ref{eq:sup:sylvestereq1}) is reduced to the equation Eq.~(\ref{eq:sup:sylvestereq2}), which is different from the one in Ref.~\cite{gong2020error}.
Using the condition (II) on the spectra of ${\bar{\mathsf H}}_{\mathsf 0\mathsf P}$ and ${\bar{\mathsf H}}_{\mathsf 0\mathsf R}$, we can adopt known relations for the Sylvester equation \cite{gong2020error,bhatia2013matrix} to obtain $\|{\bar{\mathsf T}}_{\mathsf P\mathsf R}\|\leq \|{\bar{\mathsf V}}_{\mathsf P\mathsf R}\|/\Delta_\mathsf{\mathsf P\mathsf R}$ [See also Eq.~(29) in Ref.~\cite{gong2020error}]. In addition, using a similar argument given in Eq.~(25)  in Ref.~\cite{gong2020error}, we arrive at Eq.~(\ref{eq:sup:sylvesterinequality}).

Now, we explain how we adopt the theorem to our system. 
By taking
$\hat {\mathsf H}_0 = \hat H_{\rm total} - \hat H_{\omega}$, $\hat {\mathsf V} = \hat H_{\omega}$, and $\hat{\mathsf P} = \hat P_{\Psi}=\ket{\Psi}\bra{\Psi}$ (, see Eqs.~(7) and (8) in the main text,) we have
\begin{align}
\hat {\mathsf H}_{{\mathsf P}} =
\hat P_{\Psi} \hat H \hat P_{\Psi} 
&=
(\mathbb{\hat{I}}^{\mathcal{P}}\otimes|F^{\mathcal{A} }\rangle\langle F^{\mathcal{A} }|)
\hat H
(\mathbb{\hat{I}}^{\mathcal{P}}\otimes|F^{\mathcal{A} }\rangle\langle F^{\mathcal{A} }|)\\
&=
\hat H_{\omega}^{\mathcal{P}}\otimes
|F^{\mathcal{A} }\rangle\langle F^{\mathcal{A} }|,
\end{align}
where $\mathbb{\hat{I}}^{\mathcal{P}}$ denotes the identity operator on the probe spins,
$\|\hat {\mathsf V}\| = N\omega/2$,
and $\Delta_{\mathsf P\mathsf R} \geq \min_{i}
\left[ 4\bar{J} - \sum_j |2 \Delta J_{ij}|\right]$.
Below we explain how we evaluate $\Delta_{\mathsf P\mathsf R}$.
The eigenspace $\mathcal{H}_\mathsf{R}$ in this case, which is connected to $\hat P_{\Psi}$ via the action of $\hat H_{\omega}$, is formally expressed as 
$
\mathcal{H}_\mathsf{R} :={\rm span}\{|\phi_{(k)}\rangle \otimes \left(\hat \sigma_i^x |F^{\mathcal{A} }\rangle\right)
:
|\phi_{(k)}\rangle\: (
k=1,2,…,2^{\alpha N}) $ are eigenstates of probe spins; 
$i$ is any site on ancillary spins$\}$.
In other words, $\mathcal{H}_\mathsf{R}$ is spanned by computational basis states
whose configuration includes the same configuration as $|F^{\mathcal{A} }\rangle$ except for any but one ancillary spin whose state is flipped from $|F^{\mathcal{A} }\rangle$.
Then, the energy gap between $\mathcal{H}_\mathsf{P}$ and $\mathcal{H}_\mathsf{R}$ is obtained by finding the minimum energy difference in $\hat H_{\rm total} - \hat H_{\omega}$ caused by flipping one ancillary spin in $|F^{\mathcal{A} }\rangle$.
Using the condition that every ancillary spin in $|F^{\mathcal{A} }\rangle$ is surrounded by at least three neighboring spins in the down states, we can evaluate $\Delta_{\mathsf P\mathsf R}$ as
$\Delta_{\mathsf P\mathsf R} \geq J_g :=\min_{i}
\left[ 4\bar{J} - \sum_{j\in \langle i,j \rangle} |2 \Delta J_{ij}|\right]$.
Note that $J_g$ is always positive since we assume that the spatial fluctuations of the interaction are relatively weak such that $|\Delta J_{ij}| < \bar{J}/2$.

\section{Hilbert space fragmentation with/without inhomogeneity in the couplings}\label{app:hsf}
\begin{figure}[t]
\centering
\includegraphics[width=1\columnwidth]{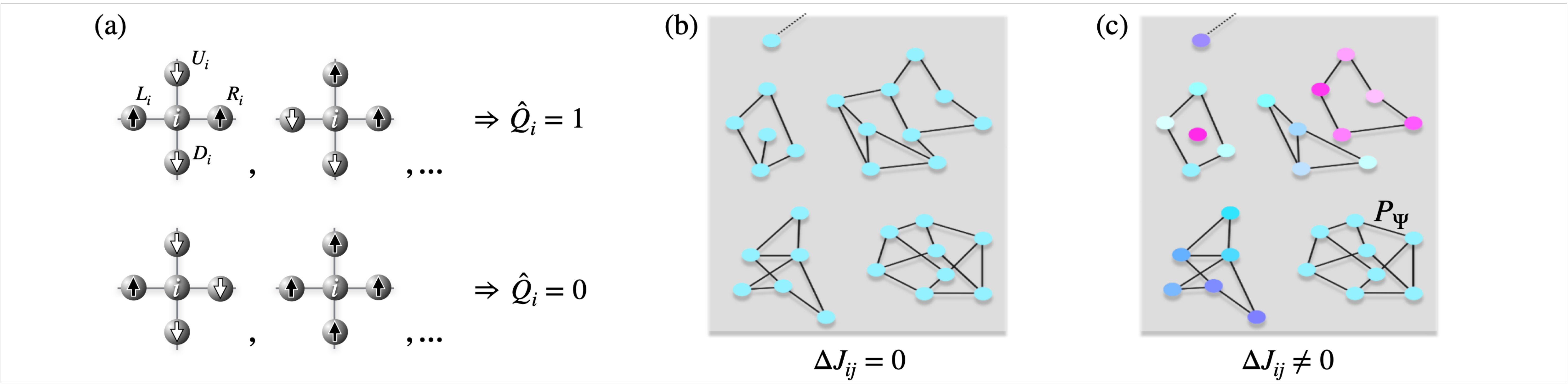}
\caption{
(a) Examples of spin configurations surrounding a site $i$ on a two-dimensional square lattice, demonstrating when the projection operator $\hat Q_i$ becomes unity or zero.
(b,c) Schematic illustration of the Hilbert space fragmentation in a single DW number sector for (b) $\hat H_{\omega}'$, i.e., the case of $\Delta J_{ij} =0$ and (c) $\hat H_{\omega}''$, i.e., the case of $\Delta J_{ij} \neq0$.
Each node represents a computational basis state having a common DW number, and the colors associated with nodes describe their expectation values for $\hat H_{\rm Ising}$ in (b) and $\hat H_{\rm Ising} - \sum_{\langle i,j \rangle} \Delta J_{ij} \hat \sigma_i^z \hat \sigma_j^z + \hat H_{\rm shift}^{\mathcal{P}}$ in (c).
Links connecting nodes exemplify nonzero off-diagonal matrix elements that the constrained Hamiltonians may possess.
Panels (b) and (c) show that the Hilbert space is still divided into parts in a similar way even after we specify the number of DWs, while the inhomogeneous case can become more complex than the homogeneous case.
Due to the additional longitudinal field $\hat H_{\rm shift}^{\mathcal{P}}$, all computational basis states in the subspace $P_{\Psi}$ have the same energy expectation value in panel (c).
}
\label{fig:sup HSF}
\end{figure}

In this section, we compare the emergent Hilbert space fragmentation (HSF) in the transverse field Ising models (TFIMs) for the cases with homogeneous ($\Delta J_{ij} = 0$) and inhomogeneous ($\Delta J_{ij} \neq 0$) Ising couplings.
As we have explained in the main text, the HSF can emerge for both cases.
We consider imposing the constraint on the Hamiltonian of the TFIM, where a transition between states due to $\hat H_{\omega}$ is suppressed when it causes the energy change much larger than $\omega$.
For the homogeneous case, the effective Hamiltonian $\hat H_{\Delta J_{ij} =0}'$, which
is obtained by  this constraint, is explicitly expressed as \cite{AmpRapiRobaClose2011,van2020quantum,yoshinaga2021emergence}
\begin{align}
\hat H_{\Delta J_{ij} =0}' = \hat H_{\omega}' + \hat H_{\rm Ising},\label{supeq:effHam}
\end{align}
with
\begin{align}
\hat H_{\rm Ising} :=& -\bar{J} \sum_{\langle i,j\rangle} \hat \sigma_i^z \hat \sigma_j^z, \label{supeq:effHamIsing}\\
\hat H_{\omega}':=& 
\frac{\omega}{2} \sum_{i} \hat \sigma_i^x \hat Q_i
\label{supeq:effHamOmega}
\end{align}
Here,
$\hat Q_{i}$ is formally given by
\begin{align}
\hat Q_i = \prod_{n = 1}^{K/2} \left((2 n)^2-\left(\sum_{j\in {\rm ngbh}(i)}{\hat\sigma_j^z} \right)^2\right)
\Big/
\left(2n \right)^2,
\end{align}
where $K$ denotes the number of the neighboring sites of a site $i$ in the lattice structure, and ${\rm ngbh}(i)$ denotes the nearest-neighbor sites of the site $i$.
Importantly, $\hat Q_{i}$ becomes an identity operator under the condition that half of the spins surrounding the $i$-th spin are in the spin-up state and the other half are in the spin-down state
in the $z$ axis, while otherwise $\hat Q_{i}$ becomes zero, as we illustrate in Fig.~\ref{fig:sup HSF}~(a) for the case of $K=4$.
This leads to emergent conservation of the domain-wall (DW) number $\hat n_{DW} := \sum_{ \langle i,j \rangle }  (1-\hat \sigma_i^z \hat \sigma_j^z)/2$, and the effective Hamiltonian is block-diagonalized.
Below, we consider the case at which the system is on a two-dimensional square lattice.

As we described in the main text, the Hamiltonian (\ref{supeq:effHam}) allows appearance of frozen regions, in which spins in the $z$ direction cannot be dynamically flipped due to the constraint \cite{yoshinaga2021emergence, hart2022hilbert}.
Hence, dynamics only occurs on non-frozen regions, i.e., melting regions, while the DW density is still conserved in each of the regions.
Indeed, each block of the effective Hamiltonian is further block-diagonalized, and these subblocks are characterised by configurations of frozen regions and DW densities on melting regions.
This means that there appear many nontrivial subspaces, where initial states in any of such subspaces cannot explore the other subspaces, i.e., the Hamiltonian shows the HSF.

A similar structure of the fragmented Hilbert space will also emerge in the Hamiltonian even if we allow for spatially inhomogeneous Ising couplings $J_{ij} = \bar{J} + \Delta J_{ij}$ with $|\Delta J_{ij}| < |\bar{J}|/2$.
In particular, the block-diagonalized structure 
again appears due to the constraint that the spin flip is energetically suppressed unless two surrounding spins are up and the other two are down, which leads to local conservation of the DW numbers.
On the other hand, the subblocks in the DW number sectors obtained for the case with $\Delta J_{ij} =0$ may further be block-diagonalized for $\Delta J_{ij} \neq0$, as we illustrate in Figs.~\ref{fig:sup HSF}~(b) and (c).
This is because a spin flip can be energetically suppressed even for the surrounding spins satisfying the two-up and two-down condition due to the energy mismatch caused by the inhomogeneous interactions.
This suggests that a spin that belongs to a melting region for homogeneous couplings may become dynamically frozen for inhomogeneous couplings, resulting in more complex fragmentation of the Hilbert space.

For completeness, we describe an effective Hamiltonian in the above as follows:
\begin{align}
\hat H_{\omega}''
:=
\sum_i
\Bigl(&
W_i^{\uparrow\uparrow\downarrow\downarrow}
{\hat P}_{L_i}^{\uparrow}{\hat P}_{D_i}^{\uparrow}\hat \sigma_i^{x}{\hat P}_{R_i}^{\downarrow}{\hat P}_{U_i}^{\downarrow}
+
W_i^{\uparrow\downarrow\uparrow\downarrow}
{\hat P}_{L_i}^{\uparrow}{\hat P}_{D_i}^{\downarrow}\hat \sigma_i^{x}{\hat P}_{R_i}^{\uparrow}{\hat P}_{U_i}^{\downarrow}
+
W_i^{\uparrow\downarrow\downarrow\uparrow}
{\hat P}_{L_i}^{\uparrow}{\hat P}_{D_i}^{\downarrow}\hat \sigma_i^{x}{\hat P}_{R_i}^{\downarrow}{\hat P}_{U_i}^{\uparrow}\nonumber\\
&+
W_i^{\downarrow\uparrow\uparrow\downarrow}
{\hat P}_{L_i}^{\downarrow}{\hat P}_{D_i}^{\uparrow}\hat \sigma_i^{x}{\hat P}_{R_i}^{\uparrow}{\hat P}_{U_i}^{\downarrow}
+
W_i^{\downarrow\uparrow\downarrow\uparrow}
{\hat P}_{L_i}^{\downarrow}{\hat P}_{D_i}^{\uparrow}\hat \sigma_i^{x}{\hat P}_{R_i}^{\downarrow}{\hat P}_{U_i}^{\uparrow}
+
W_i^{\downarrow\downarrow\uparrow\uparrow}
{\hat P}_{L_i}^{\downarrow}{\hat P}_{D_i}^{\downarrow}\hat \sigma_i^{x}{\hat P}_{R_i}^{\uparrow}{\hat P}_{U_i}^{\uparrow}
\Bigr).
\label{eq:app:effHamdef for inhomo}
\end{align}
Here, ${\hat P}_{k}^{\uparrow}:=(1+\hat \sigma_k^z)/2$ (${\hat P}_{k}^{\downarrow}:=(1-\hat \sigma_k^z)/2$) is a projection operator on the site $k$.
Also,
$L_i,D_i,R_i$, and $U_i$ denote the neighboring site left, down, right, and up to the site $i$ on the square lattice, respectively (see also Fig.~\ref{fig:sup HSF}~(a) ).
The coefficients $W_i$s take either 1 or 0 depending the energy change created by the spin flip of $i$.
Specifically, $W_i^{\uparrow\uparrow\downarrow\downarrow}$ (and the others with permutated superscripts) with $i\notin {\rm probe}$
is formally described as
\begin{align}
W_i^{\uparrow\uparrow\downarrow\downarrow}
=
W_i^{\uparrow\uparrow\downarrow\downarrow}(\delta_{\rm th})
=
\theta(\delta_{\rm th} - |\Delta J_{iL_i}+\Delta J_{iD_i}-\Delta J_{iP_i}-\Delta J_{iU_i}| ),
\label{eq:app:def wi}
\end{align}
where $\theta(x)$ is a step function which reduces to 1 (0) for $x\geq0$ ($x<0$) and $\delta_{\rm th}$ is a constant which determines the strength of the constraint.
If we set all $\Delta J_{ij}$ to zero, then all $W_i$s become 1, and the Hamiltonian (\ref{eq:app:effHamdef for inhomo}) is reduced to the Hamiltonian (\ref{supeq:effHamOmega}) with the square lattice case, see Ref.~\cite{balducci2022interface}.
On the other hand, for general $\Delta J_{ij}$, which are assumed to satisfy $\max_{i,j}2|\Delta J_{ij}|/\bar{J}<1$ as in the main text, $W_i^{\uparrow\uparrow\downarrow\downarrow}$ (and the others with permutated superscripts) can take zero. 
This is especially the case when the change in the sum of the Ising interaction energy between the site $i$ and its neighbors, which is caused by the spin flip, exceeds the threshold $\delta_{\rm th}$.
Here, to describe the physical requirement that the spin flip can occur only when the associated energy change is much smaller than $J_g$, we assume that $\delta_\mathrm{th}$ is much smaller than $J_g$ but much larger than $\omega$.
This additional suppression of some transitions, which is exemplified in Figs.~\ref{fig:sup HSF}~(b) and (c), can lead to the further fragmentation of the Hilbert space as we have explained above.
Note that, for the probe sites $i\in {\rm probe}$, we take into account the energy contribution from the shift fields $\hat H_{\rm shift}^{\mathcal{P}}$ in Eq.~(\ref{eq:app:def wi}).
As mentioned in the main text and the next paragraph, the energy difference coming from the interaction is cancelled out by the shift fields, which leads to $W_i^{\uparrow\downarrow\uparrow\downarrow}
=\theta(\delta_\mathrm{th})>0$.

Finally, we focus on the subspace $P_{\Psi}$,
which includes a state $\ket{\Psi}$,
and explain the role played by the additional longitudinal fields $\hat H_{\rm shift}^{\mathcal{P}}$ (see Eq.~(8) in the main text) in this subspace.
When we appropriately tune the bias strengths as we have shown in the main text, the energy change caused by a spin flip of each probe spin is kept zero.
In other words, the computational basis states in the subspace keep having the same energy expectation values with that for $\hat H_{\Delta J_{ij} =0}'$, see also  Fig.~\ref{fig:sup HSF}~(c).
This leads us to obtain the effective Hamiltonian $\hat H_{\omega}^{\mathcal{P}}$ (see Eq.~(7) in the main text) in this subspace.
Consequently, we can induce a coherent dynamics by constructing a certain superposition of the basis states.

\bibliographystyle{apsrev4-1}
\bibliography{BibHSFSensing}